\documentstyle[12pt]{article}
\begin{document}
\begin{center}
\textbf{THE  CASIMIR  PROBLEM  OF  SPHERICAL  DIELECTRICS: QUANTUM  STATISTICAL  AND  FIELD  THEORETICAL  APPROACHES}\\
\bigskip
\bigskip
J. S. H{\o}ye$^1$, I. Brevik$^2$, and  J. B. Aarseth$^2$ \\
\bigskip
$^1$Department of Physics, Norwegian University of Science and Technology,\\
N-7491 Trondheim, Norway\\
\bigskip
$^2$Division of Applied Mechanics, Norwegian University of Science and Technology,\\
N-7491 Trondheim, Norway
\bigskip
Revised version, February 2001\\
\end{center}
\bigskip
\bigskip

\begin{abstract}

The Casimir free energy for a system of two dielectric concentric nonmagnetic spherical bodies is calculated with use of a quantum statistical mechanical method, at arbitrary temperature. By means of this rather novel method, which turns out to be quite powerful (we have shown this to be true in other situations also), we consider first an explicit evaluation of the free energy for the static case, corresponding to zero Matsubara frequency ($n=0$). Thereafter, the time-dependent case is examined. For comparison we consider the calculation of the free energy with use of the more commonly known field theoretical method, assuming for simplicity metallic boundary surfaces. 

\end{abstract}

\newpage

\section{Introduction}

The Casimir problem for dielectrics - for general introductions see, for instance Refs.~[1-3] - has turned
out to be difficult to solve, in the presence of curved surfaces. The most typical example of a system of this sort is probably that of a single nonmagnetic compact spherical ball, surrounded by a vacuum. (Equivalently, one may imagine a spherical cavity in an otherwise uniform medium, thus dealing just with the situation typical for sonoluminescence.) Formally, in the presence of curved boundaries one is confronted with divergences when summing over all angular momenta up to infinity. This kind of divergence is usually absent when one deals with plane boundaries. Physically, the divergences are coming from the fact that phenomenological electrodynamics, implying use of the permittivity concept, becomes inappropriate at small distances. There exists a natural cutoff in the material, of the order of the intermolecular spacing, and in practice some kind of regularization has to be invoked in order to deal with the divergences in the formalism. More accurately this cutoff is the molecular diameter, as is most easily seen from the statistical mechanical method. By means of this method the electromagnetic Green function can be associated with the pair correlation function between dipole moments. The latter quantity is zero inside the hard cores, and its deviation from from the "ideal" Green function extends typically  a few molecular diameters outwards from the molecule.

In the case of nondispersive media, the use of zeta-function methods has proved to be very useful for force, or energy, calculations.  The field theory approach to the Casimir problem has been considered at various places; in addition to the references above we may mention Refs.~[4-16]. (This list is not intended to be complete; it covers mostly treatments of nonmagnetic media, and does not include the bulk of papers devoted to studies of the special case of media that satisfy the condition $\varepsilon \mu =1$. A very extensive list of references is given in the recent report of Nesterenko {\it et al.} \cite{nesterenko00a}.) The Casimir energy $E$ calculated by field theoretical methods at zero temperature for a dilute nondispersive compact sphere of radius $a$ and refractive index $n$ is positive,
\begin{equation}
E=\frac{23}{384}\frac{\hbar c}{\pi a}(n-1)^2,
\end{equation}
\label{1}
corresponding to an outward force.

Instead of making use of field theoretical methods for continuous matter, one may alternatively use {\it quantum statistical mechanical methods}. We shall in the first sections below consider methods that were developed by H{\o}ye and Stell, and others. Basic references to this kind of theory are \cite{hoye81} and \cite{hoye82}. In the Casimir context, H{\o}ye and Brevik \cite{hoye98} used the quantum statistical mechanical path integral method to calculate the van der Waals force between dielectric plane plates. Recently, we have applied the same method to a single compact spherical ball \cite{hoye00}. This statistical method, although probably not so well known as the field theoretical methods, turns out to be quite powerful. Thus, we can use it to calculate explicitly the short range terms in the single sphere's free energy, and verify how the repulsive Casimir surface force as calculated by field theoretical methods is simply a residual, cutoff independent, term in a complicated expression containing many terms. Cf. in this context also Refs.~\cite{barton99}, \cite{barton00}, and \cite{bordag99}.  

It is now natural to ask: what is the experimental status in this field? Recently, there has been an impressive improvement of the experimental accuracy as regards force measurements; one has been able to verify the theoretically predicted Casimir forces, lying in the piconewton range, up to an accuracy of about 1 per cent. In Ref.~\cite{lamoreux97} the Casimir force was demonstrated between metallic surfaces of a sphere above a disk using a torsion pendulum, whereas in Refs.~\cite{mohideen98}, \cite{roy99} an atomic force microscope was used.

One important lesson is to be learned from these experimental works is the following: they alway involve {\it two} (in principle there may even be more) bodies. The Casimir surface force on a {\it single} sphere is not measured. There seems not even to be an idea of how to measure such a force; probably this reflects simply that the force concept as such is not observationally well defined. Thus, in order to keep contact with experiments, at least in principle, one ought to consider at least two bodies. And this brings us to the theme of the present paper, namely to calculate the mutual free energy for a system of two spherically-shaped concentric nonmagnetic dielectrics. We will envisage that there is one compact sphere for $r<a$, and one semi-infinite similar medium for $r>b$, so that there is a vacuum gap of width $d=b-a$ in between. There will be an attractive Casimir force between the two media. One may object that there is still no straightforward way to imagine measuring such a force; however this does not create difficulties for our main purpose, which is to calculate the Casimir force in a setting which maintains spherical symmetry and yet avoids the complications with internal, cutoff dependent, forces.

In the following four sections we shall deal with the quantum statistical mechanical theory, with an emphasis on the static limit (zero Matsubara frequency, where derivations are more simple). Thereafter, for comparison we consider the alternative field theoretical approach, limiting us for simplicity to the case of perfect metallic walls at $r=a,b$. The resulting expressions for the free energy are Eq.~(18) for the static case (Matsubara frequency equal to zero) and Eq.~(40) for the time-dependent case (general Matsubara frequency). These expressions are obtained within the statistical mechanical approach. Within the field theoretical approach, the finite-temperature free energy is given by Eq.~(68) assuming, as mentioned, perfectly conducting walls.

There is one notable difference between the statistical mechanical approach and the field theoretical approach, as far as the free energy is concerned. In the first of these cases the method is basically more simple, at least in principle, as one needs only knowledge about the mode eigenvalues of the oscillating dipole moments in the dielectric medium; cf. Eq.~(3). These eigenvalues are again related to the pair correlation function. In the second case the calculation is more indirect, as one first calculates the surface force (arising from the mutual interaction) on the outer surface implying use of Maxwell's stress tensor, and thereafter relates the force to the free energy via integration of Eq.~(67). That is, the field theoretical method involves use of the two-point functions for the electric and magnetic fields. As we also want to show that the results obtained by these widely different methods are in agreement, we consider some cases that are easy to analyse analytically, by both methods.

We ought to stress again the conceptual difference between the two methods studied in this paper. By the field theoretical method the Casimir effect is regarded as the energy shift due to the frequency eigenvalues of the quantized electromagnetic field in the presence of dielectric media. By the more recent statistical mechanical method this energy shift is regarded as a consequence of the {\it dipole-dipole interaction} between oscillating dipole moments embedded in the molecules of the media. The latter viewpoint can be realized by exploiting the analogue that exists between phonons in a solid, and electromagnetic waves or photons in vacuum. Then impurities in the solid will be the analogue to dielectric particles. These impurities will couple to the phonons of the solid and modify their frequencies. If, however, one wants to do the statistical mechanics of the latter system, e.~g. to calculate its free energy, then one can first integrate out the coordinates of the pure solid that appear in the path integral representation of the quantized problem. With harmonic oscillators one encounters in this way Gaussian integrals which can easily be calculated. The result is independent of the impurities, except that {\it interactions} between them are introduced. Thus, the resulting change in free energy can be related to these induced induced interactions (being in general time-dependent) in the system of impurities. Likewise, in this picture the dipole-dipole interactions are related to the Casimir free energy.

We employ Gaussian electromagnetic units in this paper.

\section{General remarks}

Consider the free energy $F(T)$ due to the mutual interaction between two spherical dielectric bodies with concentric surfaces at $r=a$ and $r=b$. The attractive Casimir force between the surfaces, per unit area at the outer surface, is equal to $f=-1/(4\pi b^2)\partial F/\partial b$. As shown earlier for the case of plates \cite{hoye98}, this Casimir force can be interpreted as the dispersion force arising from thermal fluctuations of molecular dipole moments. By our quantum statistical mechanical considerations this also incorporates the quantum fluctuations at $T=0$. That means, all fluctuations can be regarded as purely thermal for any $T$. The difference between classical and quantum situations is that in the former case these fluctuations vanish at $T=0$ while in the latter case they remain finite. In \cite{hoye00} we considered the low density (or small  $\varepsilon -1$) version of the single-body problem showing, as mentioned above, that the divergences are due to the continuum model of the medium. A cutoff in length scale is needed. For realistic systems this cutoff is determined by the molecular hard cores through their influence upon the pair correlation function. For two polarizable particles the free energy due to their mutual attraction can be written as \cite{{hoye98},{brevik88}}
\begin{equation}
\beta F= -\frac{1}{2}\sum_{n=1}^{\infty} \frac{1}{n} (\alpha_1 \psi \alpha_2 \psi)^n=
\frac{1}{2}\ln (1-\alpha_1\psi\alpha_2\psi),
\end{equation}
\label{2}
with  $\beta=1/k_B T$. The $\psi$ represents the potential energy of the dipole-dipole interaction which for general $K$ (see Eq.~(3) below) is given by Eqs.~(6) and (7) below. 
Now the two particles can be generalized to and regarded to be our two spherical bodies, in the same way as two semi-infinite parallel plates were treated in \cite{hoye98}. That means, the two spherical bodies are regarded as two particles with many internal degrees of freedom. In this way Eq.~(2) becomes a short hand notation wherein $\psi$ represents the interaction between two points in the two bodies (over which we integrate), and the polarizabilities $\alpha_1$ and $\alpha_2$ become the respective internal correlation functions of the two bodies with their mutual interaction $\psi$ switched off. As noted in \cite{hoye98} the expression (2) is formally exact for coupled harmonic oscillators, i.e., the model that we are employing for the polarizable particles. In terms of graphs, the expression (2) represents the ring graphs in the $\gamma-$ordering for the long-range forces, $\gamma$ being the inverse range of interaction \cite{hemmer64}. For coupled oscillators Eq.~(2) above and Eq.~(3) below are exact results \cite{hoye80}.

Extending to the quantum mechanical case, Eq.~(2) generalizes to
\begin{equation}
\beta F=\frac{1}{2}\sum_K \ln(1-\alpha_{1K}\psi_K\alpha_{2K}\psi_K),
\end{equation}
\label{3}
where $K=2\pi n/\beta$  with $n$ integer (i.e., $n \in \langle -\infty, \infty \rangle )$. Note that $K=\hbar \zeta_n$, where $\zeta_n$ is the Matsubara frequency $\zeta_n=-i\omega$ and $\omega$ is the frequency.

For low density (or small $\alpha$) only the first term in the sum (2) is needed. This is the situation considered in \cite{hoye00} and found there, after some transformations, to be in agreement with earlier works. The first term means simply that one takes the (radiating) dipole interaction squared, average (integrate) over the fluctuating dipole moments, and finally integrate over the two media. Thus
\begin{equation}
F=\rho^2\int d{\bf r}_1d{\bf r}_2\,\Phi,
\end{equation}
\label{4}
where $\rho$ is the particle density,  $ r_1<a,~~r_2>b$, and  \cite{hoye98}
\begin{equation}
\beta \Phi=-\frac{3}{2}\sum_K\alpha_K^2\left[ 2\psi_{DK}^2(r)+\psi_{\Delta K}^2(r)\right],
\end{equation}
\label{5} 
where $ {\bf r}={\bf r}_2-{\bf r}_1$.
(Here ${\bf r}_1$ and ${\bf r}_2$ are positions in different media, so double counting does not occur.) The radiating dipole interactions used in (5) can be written as
\begin{equation}
\psi(12)=\psi_{DK}(r)D_K(12)+\psi_{\Delta K}(r)\Delta_K(12),
\end{equation}
\label{6}
with 
\begin{eqnarray}
D_K(12)           &=& 3(\hat{r}\,\hat{a}_{1K})(\hat{r}\,\hat{a}_{2K})-\hat{a}_{1K}\,\hat{a}_{2K}, \nonumber  \\    \Delta_K(12)      &=& \hat{a}_{1K}\hat{a}_{2K}. \nonumber
\end{eqnarray}
Here the hats denote unit vectors, and ${\bf a}_{iK}$ is the Fourier transform of the fluctuating dipole moment of particle number $i$ in imaginary time; cf. Eq.~(5.2) in \cite{brevik88}. Explicitly, from Eq.~(5.10) in \cite{brevik88},
\[ \psi_{DK}(r)=-\frac{e^{-\tau}}{r^3}\left( 1+\tau+\frac{1}{3}\tau^2 \right), \]
\begin{equation}
\psi_{\Delta K}(r)=-\frac{e^{-\tau}}{r^3}\frac{2}{3}\tau^2+\frac{4\pi}{3}\delta({\bf r}),
\end{equation}
\label{7}
with
\[ \tau=\frac{i\omega r}{c}=-\frac{Kr}{\hbar c}~~~{\rm for}~~~ \Im (\omega)<0~~(\mbox{or}~K<0), \]
and $-K \rightarrow |K|$  when extending to $K>0$ in (7) (see Eq.~(5.11) in \cite{brevik88}). 

For general $K$ we are not able to calculate the integral (4) in a direct way (but we can calculate it indirectly for arbitrary density, as will be argued later, in Sect.~4). We can integrate, however, for $K=0$. The integral of interest then becomes
\begin{equation}
I=\int_{r_1<a,\, r_2>b} d{\bf r}_1d{\bf r}_2\frac{1}{r^6},~~~r^2=r_1^2+r_2^2-2r_1r_2\cos\theta.
\end{equation}
\label{8}
This integral can be evaluated in closed form \cite{milton97} (second reference). However, in Sect.~3 we want to extend this low density evaluation to the case of arbitrary density or arbitrary $\varepsilon$. To do so, we need the contributions related to the various spherical harmonics. Thus we will here perform an expansion of the integral. This will also be used as an independent verification at the end of Sect.~4, where the more general theory developed there for arbitrary values of $K$ turns out to yield the correct result when $K \rightarrow 0$.

Using spherical coordinates to integrate over the angle $\theta$ between ${\bf r}_1$ and ${\bf r}_2$ we first obtain ($x=-\cos \theta$)
\begin{eqnarray}
J=\int_{-1}^{1}\frac{dx}{(r_1^2+r_2^2+2r_1r_2x)^3}
&=& \frac{1}{4r_1r_2}\left[\frac{1}{(r_2-r_1)^4}-\frac{1}{(r_2+r_1)^4} \right] \nonumber \\
&=& \frac{1}{2r_1r_2^5}\sum_{l=1}^\infty \frac{(2l+2)(2l+1)2l}{6}\left(\frac{r_1}{r_2}\right)^{2l-1},
\end{eqnarray}
\label{9}
performing a series expansion to make it easy to relate to the result for high density. Then,
\begin{equation}
I=8\pi^2\int_0^a r_1^2dr_1\int_b^\infty r_2^2dr_2J=\frac{8\pi^2}{3}\sum_{l=1}^\infty
\frac{(l+1)l}{2l+1}\sigma_l,
\end{equation}
\label{10}
with $\sigma_l=(a/b)^{2l+1}$.

Inserted in (4) we obtain ($\alpha_0=\alpha$)
\begin{equation}
\beta F=\beta \rho^2\frac{3}{2}\alpha^2\cdot 2\cdot I=-\frac{1}{2}(\varepsilon-1)^2
\sum_{l=1}^\infty \frac{(l+1)l}{2l+1}\sigma_l.
\end{equation}
\label{11}
Note that for small $\varepsilon-1$, Eq.~(11) will be the high temperature result for which only $K=0$ contributes. This high temperature result at low density may in itself be of limited interest as it does not go beyond earlier results. But here we use it as a basis to make further developments. So in the next section the formalism is generalized to arbitrary density or $\varepsilon$, although it is still restricted to $K = 0$. In Sect.~4, a derivation that encompasses both arbitrary $\varepsilon$ and $K$ is given.

\section{The static case}

For simplicity we first consider the {\it static} case, by which we mean that the frequency is zero ($K=0$). Then the electromagnetic dipole-dipole interaction is the well-known static, time-independent (also called  instantaneous) one. By the {\it time-dependent} case we mean the general situation with $K \neq 0$. Then the dipole-dipole interaction will be the radiating or dynamical electromagnetic field where the time delay due to the finite speed of light is involved. Note that in general both the static and the dynamic cases contribute to the Casimir effect. The former, being proportional to $T$, contains the whole effect when $T \rightarrow \infty$, but vanishes when $T \rightarrow 0$.

For general $\varepsilon$ one should sum up the series in Eq.~(2). This will not be a simple task. However, one can include a strength factor $\lambda$ along with the perturbing interaction $\psi$ and differentiate (2) (or (3)) to obtain
\begin{equation}
{\beta \frac{\partial F}{\partial \lambda}}|_{\lambda=1}=-\psi(\alpha_1\psi_c\alpha_2),
\end{equation}
\label{12}
where
\[ \psi_c=\frac{\psi}{1-\alpha_1\psi\alpha_2\psi}. \]
Here the $\alpha_1\psi_c\alpha_2$ will be the pair correlation function for the fluctuating dipole moments. As shown in Appendix A in \cite{hoye98} the $\psi_c$ (apart from a simple factor) is the Green function for the electromagnetic problem with the dielectric medium present while $\psi$ is the one for vacuum. Thus we can utilize Maxwell's equations for electrostatics to obtain this zero frequency Green function or $\psi_c$ in the presence of two dielectric spheres.

The electrostatic potential $\Phi$ fulfils the Laplace equation $\nabla^2 \Phi=0$ with $\varepsilon =$const. Splitting off the spherical harmonic factor $Y_{lm}=Y_{lm}(\theta, \varphi)$,
\begin{equation}
\Phi=\Phi_l(r)Y_{lm}(\theta, \varphi),
\end{equation}
\label{13}
we can write the radially dependent term in the form
\begin{eqnarray}
\Phi=\left\{ \begin{array}{ll}
\frac{1}{\varepsilon}\left(\frac{a}{r} \right)^{l+1}+B\left(\frac{r}{a}\right)^l,&  r<a \\
C\left(\frac{a}{r}\right)^{l+1}+C_1\left(\frac{r}{a}\right)^l,                   & a<r<b \\
D\left(\frac{a}{r}\right)^{l+1},                                                 & b<r.
\end{array}
\right.
\end{eqnarray} 
\label{14}
From the boundary conditions the coefficients can be determined. We give the coefficient $D$ belonging to  the exterior region
\begin{equation}
D=\frac{(2l+1)^2}{(\varepsilon-1)^2(l+1)l}\frac{A_l}{(1-A_l\sigma_l)},
\end{equation}
\label{15}
where
\begin{equation}
 A_l=\frac{(\varepsilon-1)^2(l+1)l}{[\varepsilon(l+1)+l](\varepsilon l+l+1)}~~~~
\mbox{and}~~~\sigma_l=\left(\frac{a}{b}\right)^{2l+1}. 
\end{equation}
\label{16}
The coefficient $D$ represents the change of the field for $r>b$ relative to the $\varepsilon=1$ case for a given point source. Via Eq.~(12) the free energy is now obtained in a straightforward way. For small $\varepsilon -1 $ the quantity (12) is twice the quantity (11). Thus the expression (12) becomes
\begin{equation}
\beta \frac{\partial F}{\partial \lambda} |_{\lambda=1}=-\sum_{l=1}^{\infty} (2l+1)
\frac{A_l\sigma_l}{1-A_l\sigma_l}.
\end{equation}
\label{17}
As we will argue below $A_l\sigma_l$ will be proportional to the strength factor squared, $\lambda^2$. Thus integrating (17) we obtain the free energy
\begin{equation}
\beta F=\frac{1}{2}\sum_{l=1}^\infty (2l+1)\ln(1-A_l\sigma_l),
\end{equation}
\label{18}
which clearly yields an attractive force between the two spherical bodies.

It should be noted that the series (18) is convergent as it contains no self-energy. From Eq.~(16) it is seen that $A_l \leq 1 $ and $\sigma _l $ decays like an exponential as $l \rightarrow \infty$, implying $A_l \sigma_l <1$. Further, no cutoff is needed as Eq.~(18) only includes the free energy shift due to the mutual interaction between the two bodies. A "self-energy" would arise if interactions within each body were considered. But yet the necessary minimum distance between molecules would prevent the latter expression from diverging \cite{hoye00}.

\section{Further analysis of the static case}

The time-dependent case ($K \neq 0$) will be more complex to handle as we are not able to perform analytically the generalization of the integration (9) that gave the results (11) and (18). We find however that this case can be handled indirectly, noting that the quantity
\begin{equation}
M=\alpha_1\psi\alpha_2\psi
\end{equation}
\label{19}
in Eqs.~(2) and (12) can be regarded as a matrix. We want the trace of these expressions (as well as the expression (3)), which amounts to integrating over positions and dipolar moments of the particles. The matrix can be transformed into a diagonal matrix $\Lambda$ through some matrix $S$,
\begin{equation}
M=S\Lambda S^{-1}.
\end{equation}
\label{20}
Then
 \[   Tr(M^q)=Tr(\Lambda^q)=\sum_i\lambda_i^q,  \]
where
$\lambda_i $ are the diagonal elements of $\Lambda$. Also,
\begin{equation}
(M^q)_{ij}=\sum_l S_{il}(S^{-1})_{lj}\lambda_l^q.
\end{equation}
\label{21}
Thus to obtain the free energy (2) or (3) we only need the eigenvalues $\lambda_l$. Use of the spherical harmonics $Y_{lm}$ for our present problem produces such a diagonalization and, as the results (17) and (18) show, the $A_l\sigma_l$ represent these eigenvalues. The prefactor $2l+1$ is simply the degeneracy factor.

However, without performing the integration (10) the identification of $A_l\sigma_l$ with the appropriate eigenvalues is not obvious and cannot be concluded from Eq.~(15) alone.  Then we turn to Eq.~(12) and consider the correlation function (or the equivalent Green function) which can be expanded as
\begin{equation}
\alpha_1\psi_c \alpha_2=\frac{\alpha_1 \psi \alpha_2}{1-\alpha_1\psi\alpha_2\psi}=
\alpha_1\psi\alpha_2 \sum_{n=0}^{\infty}M^n
\end{equation}
\label{22}
with $M=\alpha_1\psi\alpha_2\psi $. Applying $S$ the $M$ is made diagonal such that  
\begin{equation}
S^{-1}\,\alpha_1\psi_c\alpha_2\,S=S^{-1}\,\alpha_1\psi\alpha_2\,S\frac{1}{1-\Lambda},
\end{equation}
\label{23}
and $\Lambda$ or its eigenvalues $\lambda_i$ can be identified via the ratio
\begin{equation}
S^{-1}\,\frac{\alpha_1\psi_c\alpha_2}{\alpha_1\psi\alpha_2}\,S=\frac{1}{1-\Lambda}.
\end{equation}
\label{24}
Here the $D$ in Eq.~(15) represents the numerator (the full correlation function), while the denominator can be identified with the first term in a chain bond expansion with one single potential bond $\psi$ and two hypervertices $\alpha_1$ and $\alpha_2$ (or correlation functions for the two media with their mutual interaction switched off). By chain bond expansion we mean the graphical representation of the terms in the expansion (22), where hypervertices $\alpha_i$ alternate with potential bonds $\psi$. This notation has its origin in the statistical mechanical theory of fluids \cite{hemmer64}. To go beyond standard mean field theory the Mayer graph expansion can be rearranged, and the chain bond will then become the leading correction (for forces of long range) to the correlation function of the reference system (e.g., hard spheres). In the present case with harmonic oscillators the expansion  turns out to be exact for the pair correlations of amplitudes (corresponding to Gaussian fluctuations).

The first term of the expansion (22) is now obtained by considering the two spherical bodies separately, or equivalently by considering Eq.~(14) first with $a=0$ and thereafter with $b=\infty$. Then there will be no multiple $\psi $ bonds going back and forth, as there are no longer two media present. First take away the inner sphere, which is done by putting $a=0$. Solving for $D$ one obtains ($B=0$)
\begin{equation}
D= D_0=\frac{2l+1}{\varepsilon(l+1)+l}C.
\end{equation}
\label{25}
The amplitude ratio $D_0/C$ will represent $\alpha_2$ (or $\alpha_1$). Secondly, take away the outer sphere, which is done by putting $b=\infty$. Solving for $C$ one then obtains ($D=0, ~C_1=0$)
\begin{equation}
C= C_\infty =\frac{2l+1}{\varepsilon l+l+1}.
\end{equation}
\label{26}
The amplitude ratio $C_\infty/(1/\varepsilon)$ (see Eq.~(14)) will represent $\alpha_1$ (or $\alpha_2$). Thus $D_0$ will represent $\alpha_1 \psi \alpha_2$ as $\psi$ is represented by $1/\varepsilon$. Likewise the $D$ as given by Eq.~(15) represents the full correlation (or Green) function as both spheres are present, by which the ratio $D/D_0$ yields the sought eigenvalues of Eq.~(24).

Combining Eqs.~(25) and (26),
\begin{equation}
D_0=\frac{(2l+1)^2}{[\varepsilon(l+1)+l](\varepsilon l+l+1)}.
\end{equation}
\label{27}
Relating this to Eqs.~(15) and (24) we see that the eigenvalues of $\Lambda$ are $\lambda_l=A_l\sigma_l,~D/D_0=1/(1-\lambda_l)$. Thus we recover the results (17) and (18) when $\lambda_l$ is used in the expressions (12) and (2), taking into account the degeneracy factor $2l+1$. That means, the present indirect approach is fully consistent with the explicit integration (10) that led to the same results.

\section{The time-dependent case}

Including the time dependence the solutions of Maxwell's equations become more complex. One has to solve the vector wave equation, and the fields have to satisfy the boundary conditions at the two surfaces. Again the spherical harmonics $Y_{lm}$ can be used, and the remaining problem how to decompose the vector fields parallel and transverse to the spherical surfaces is conveniently dealt with in terms of the TM (transverse magnetic) and TE (transverse electric) mode \cite{jackson75}. Application of the angular momentum operator $ {\bf L}=(1/i){\bf r \times \nabla}$  (with $\hbar=1$) creates a vector normal to ${\bf r}$, i.~e.  ${\bf r\cdot L}=0 $, and is thus parallel to the spherical surfaces. As ${\bf L}$ commutes with the $\nabla^2$ operator of the wave equation, and as ${\bf L}$ does not contain differentiation with respect to $r$, the wave equation has TM solutions of the form
\begin{equation}
{\bf B}=\frac{\Phi(r)}{r}{\bf L}Y_{lm},
\end{equation}
\label{28}
where $\Phi(r)$ is some function of $r$. Likewise the TE solutions follow with ${\bf B}$ replaced by ${\bf E}$. 

The ${\bf E}$ field is now obtained from
\begin{equation}
{\bf \nabla \times B}=\frac{\varepsilon}{c}\frac{\partial {\bf E}}{\partial t}=-ik\varepsilon {\bf E},
\end{equation}
\label{29}
where $k=\omega /c$. Thus we need the formula \cite{jackson75}
\begin{equation}
i{\bf \nabla\times L}={\bf r}\nabla^2-{\bf \nabla}(1+r\frac{\partial}{\partial r}).
\end{equation}
\label{30}
Now applying boundary conditions on the spherical surfaces, we find that the condition on the radial component of ${\bf E}$ coincides with that of ${\bf B}$, so that we need its component ${\bf E}_\perp$ transverse to ${\bf r}$ which comes from the last term in (30) where only derivatives with respect to the polar angles are needed from the ${\bf \nabla}$ operator. The latter again act only on $Y_{lm}$ which are the same on both sides of the interfaces and can thus be disregarded as far as boundary conditions are concerned. Therefore we are left with the $r$-dependence of ${\bf E}$ , where the term of interest is given by
\begin{equation}
( 1+r\frac{d}{dr} ) ( \frac{\Phi(r)}{r}) = \frac{d\Phi(r)}{dr}.
\end{equation}
\label{31}
The solutions of the wave equation for a given frequency are the Riccati-Bessel functions. As independent pair of functions it is convenient to choose the functions that are proportional to $rj_l(kr)$ and to $rh_l^{(1)}(kr)$; the first one because of its finiteness at the origin, the second because of outgoing boundary conditions at infinity. After frequency rotation, and convenient normalization, these are the functions denoted by $s_l$ and $e_l$ in the field theory section below, in Eq.~(53). For simplicity we will in the present section omit the subscript $l$. We will let  subscripts $a,b$ refer to functions taken at $r=a,b$, and  add an extra subscript $\varepsilon$ to indicate that the function is taken inside a dielectric medium. Like Eq.~(14) we can now write
\begin{eqnarray}
\Phi(r)=\left\{ \begin{array}{ll}
e_\varepsilon +Bs_\varepsilon,  & r<a \\
Ce+C_1s,              & a<r<b \\
De_\varepsilon,       & b<r.
\end{array}
\right.
\end{eqnarray}
\label{32}
As compared to Eq.~(14) the coefficient $1/\varepsilon$ for $r<a$ has been dropped since the $\Phi(r)$ represents the magnetic field, but has no further consequence as it only affects the other coefficients by a proportionality factor $\varepsilon$. Requiring continuity of the tangential components ${\bf B}_\perp$ and ${\bf E}_\perp$ across the surfaces we obtain the equations
\[ e_{a\varepsilon}+Bs_{a\varepsilon}=Ce_a+C_1s_a, \]
\[ \frac{1}{\varepsilon}(e_{a\varepsilon}'+Bs_{a\varepsilon}')=Ce_a'+C_1 s_a', \]
\[ Ce_b+C_1 s_b=De_{b\varepsilon}, \]
\begin{equation}
Ce_b'+C_1s_b'=\frac{D}{\varepsilon}e_{b\varepsilon}',  
\end{equation}
\label{33}
where we emphasize that the primes here mean differentiation with respect to $r$.

To obtain the eigenvalues of interest we now proceed as in Sect. 4. So like Eq.~(25) we find from the two last members of Eq.~(33) (i.~e.,  $a=0$)
\begin{equation}
D= D_0=c_2C, ~~~~\mbox{with}~~~~c_2=\varepsilon\frac{e_bs_b'-e_b's_b}{\varepsilon e_{b\varepsilon}s_b'-
e_{b\varepsilon}'s_b},
\end{equation}
\label{34}
and like Eq.~(26) we find from the two first members of Eq.~(33) (i.~e., $b=\infty$ and $C_1=0$)
\begin{equation}
C=C_\infty =c_1,~~~~\mbox{with}~~~~c_1=\frac{e_{a\varepsilon}'s_{a\epsilon}-e_{a\varepsilon}s_{a\varepsilon}'}
{\varepsilon e_a's_{a\varepsilon}-e_as_{a\varepsilon}'}.
\end{equation}
\label{35}
Combining these we obtain, like Eq.~(27),
\begin{equation}
D_0=c_1c_2.
\end{equation}
\label{36} 
As explained below Eq.~(24) and in connection with Eqs.~(25) and (26), this $D_0$ represents the chain with a single potential bond $\psi$. The full chain bond (see explanation below Eq.¨(24)) will be obtained by solving Eqs.~(33) as they stand. This yields 
\begin{equation}
D=\frac{D_0}{1-\lambda_{\varepsilon l}},
\end{equation}
\label{37}
where
\begin{equation}
\lambda_{\varepsilon l}=\frac{(\varepsilon s_a's_{a\varepsilon}-s_a s_{a\varepsilon}')
(\varepsilon e_b'e_{b\varepsilon}-e_b e_{b\varepsilon}')}
{(\varepsilon e_a's_{a\varepsilon}-e_a s_{a\varepsilon}')
(\varepsilon e_{b\varepsilon}s_b'-e_{b\varepsilon}'s_b)}
\end{equation}
\label{38}
are the eigenvalues of interest in the construction of the free energy; cf. the argument above Eq.~(27) in Sect. 4.

The static case ($\omega=0$) is recovered by putting $e_\varepsilon=e=1/r^l$ and $s_\varepsilon=s=r^{l+1}$, which yields $\lambda_{\varepsilon l}=A_l\sigma_l$ in accordance with Eq.~(15). Note that Eqs.~(33) are somewhat different from those used in the static case as $\varepsilon$ is replaced by $1/\varepsilon$ while $l$ and $l+1$ are interchanged, but the result is the same.

When $\omega \neq 0$ there is also another set of modes, namely the TE modes. They are obtained by replacing ${\bf B}$ with ${\bf E}$ in Eq.~(28), and by interchanging ${\bf B}$ and ${\bf E}$ in Eq.~(29) while removing the factor $\varepsilon$ and the minus sign on the right hand side. Again imposing boundary conditions, Eqs.~(33) are recovered, except that the factor $1/\varepsilon$ is no longer present. Solving for $D$ we recover the results (34)-(38) also, except that all factors  $\varepsilon$ are no longer present. The eigenvalues of interest now become
\begin{equation}
\lambda_l=\frac{(s_a's_{a\varepsilon}-s_a s_{a\varepsilon}')
(e_b' e_{b\varepsilon}-e_b e_{b\varepsilon}')}
{(e_a' s_{a\varepsilon}-e_a s_{a\varepsilon}')
(e_{b\varepsilon}s_b'- e_{b\varepsilon}'s_b)}.
\end{equation}
\label{39}
With the eigenvalues (38) and (39), the expression (18) for the free energy can be extended in a straightforward way to the time-dependent case, and we get
\begin{equation}
\beta F=\frac{1}{2}\sum_K \sum_{l=1}^{\infty}(2l+1)[\ln (1-\lambda_{\epsilon l})+\ln (1-\lambda_l)],
\end{equation}
\label{40}
where the prefactor $2l+1$ is again the degeneracy factor, and $K=2\pi n/\beta$ with $n$ integer (i.e., $n\in \langle -\infty, \infty \rangle$).

Finally it can be noted that for two parallel plates separated by a distance $d$ the free energy can be written in a similar general form. This energy can be found by integrating the surface force given by Eq.~(2.9) of Ref. \cite{hoye98}. This surface force is the famous Lifshitz result \cite{lifshitz56}. The parallel plates result for the mutual free energy per unit area becomes 
\begin{equation}
\beta F= \frac{1}{4\pi}\sum_K \, \int_{\zeta_n/c}^\infty \left[ \ln (1-\lambda_{\varepsilon q})+
\ln (1-\lambda_q) \right] q\, dq, 
\end{equation}
\label{41}
where
\[ \lambda_{\varepsilon q}=A_n e^{-2qd},~~~~\lambda_q=B_n e^{ -2qd}, \]
with
\[ A_n=\left( \frac{\varepsilon -\kappa}{\varepsilon +\kappa} \right)^2,~~~
B_n=\left( \frac{1-\kappa}{1+\kappa}\right)^2, \]
\[ \kappa=\left[ 1-(\varepsilon -1)\left(\frac{\omega}{cq}\right)^2 \right]^{1/2},
~~~K=-i\hbar \omega=\hbar \zeta_n=2\pi n/\beta. \]
It should be noted that Eqs.~(40) and (41) are general results, valid for arbitrary permittivity $\varepsilon (\omega)$ and temperature, and they give the free energy due to the mutual interaction between the two media. There are no diverging self-energy terms. From Eq.~(37) one must expect $\lambda_{\varepsilon l},~\lambda_l < 1$, and in the $K=0$ case (18) one has the converging factor $\sigma_l=(a/b)^{2l+1}$. This convergence will not be weakened when $K \neq 0$ as then exponential factors also enter the solution of the wave equation for imaginary frequency.

\section{Field-theoretical approach: the surface force}

We now consider, as an alternative, the field theoretical approach to the same physical system, with the simplification, however, that the compact media are perfect conductors ($\varepsilon=\infty $). Thus we can compare with the above general result in this special case. We shall make use of the local Green-function method, as developed in particular by Schwinger and his school. A basic reference to this kind of theory applied to the case of spherical symmetry (a perfecly conducting shell) is Milton {\it et al.} \cite{milton78}. To our knowledge Milton was also the first to apply this theory to the compact ball problem \cite{milton80}. Generalization of the theory, so as to take into account electrostriction, was made by Brevik \cite{brevik82}. Later references are \cite{brevik94}-\cite{papers99} and \cite{{brevik99}, {brevik99a}}. (This list does not include the main part of the references dealing with $\varepsilon\mu=1$ media, as well as papers dealing with the mode summation method.) We now put $\hbar =c=k_B=1$.

Once the assumption about perfect conductors is accepted, the formalism becomes relatively simple. Since all fields in the regions $r<a$ and $r>b$ are equal to zero, we have to consider the fields in the vacuum gap only. The Green function ${\bf \Gamma}(x,x')$ for two spacetime points $x$ and $x'$ has a Fourier transform ${\bf \Gamma}({\bf r},{\bf r'},\omega)$ defined by
\begin{equation}
{\bf \Gamma}(x,x')=\int_{-\infty}^{\infty}\frac{d\omega}{2\pi}e^{-i\omega\tau}\,{\bf \Gamma}({\bf r},{\bf r'}, \omega),
\end{equation}
\label{42}
with $\tau=t-t'$. Note that the convention of Fourier transform used here implies a change of sign of $\omega$ (i.~e., $\omega \rightarrow -\omega$), compared to the definition used in the preceding sections, e.~g. Eq.~(7). The governing equation for ${\bf \Gamma}$, as following from Maxwell's equations, is
\begin{equation}
{\bf \nabla\times\nabla\times\Gamma(r,r'},\omega)-\omega^2{\bf \Gamma(r,r'},\omega)=\omega^2{\bf 1}\delta({\bf r-r'}),
\end{equation}
\label{43}
and the spectral two-point function for the electric field components is
\begin{equation}
i\langle E_i({\bf r})E_k({\bf r'})\rangle_{\omega}= (4\pi)\Gamma_{ik}({\bf r,r'},\omega)
\end{equation}
\label{44}
(the prefactor $4\pi$ appearing because of our present use of the Gaussian system of units). There are two scalar Green functions, $F_l(r,r')$ and $G_l(r,r')$, since there are two independent field modes. The connection between these functions and the spectral two-point functions is
\begin{equation}
i\langle E_r(r)E_r(r')\rangle_\omega=\frac{(4\pi)}{rr'}\sum_{l=1}^\infty \frac{2l+1}{4\pi}
l(l+1)G_l(r,r'),
\end{equation}
\label{45}
\begin{equation}
i\langle E_\perp (r)E_\perp(r')\rangle_\omega=(4\pi)\sum_{l=1}^\infty \frac{2l+1}{4\pi}
[ \omega^2F_l(r,r')+\frac{1}{rr'}\frac{\partial}{\partial r}r\frac{\partial}{\partial r'}r'
G_l(r,r') ],
\end{equation}
\label{46}
\begin{equation}
i\langle H_r(r)H_r(r')\rangle_\omega=\frac{(4\pi)}{rr'}\sum_{l=1}^\infty\frac{2l+1}{4\pi}
l(l+1)F_l(r,r'),
\end{equation}
\label{47}
\begin{equation}
i\langle H_\perp (r)H_\perp (r')\rangle_\omega=(4\pi)\sum_{l=1}^\infty
[\omega^2G_l(r,r')+\frac{1}{rr'}\frac{\partial}{\partial r}r\frac{\partial}{\partial r'}r'F_l(r,r')].
\end{equation}
\label{48}
Here, it is assumed that the vectors ${\bf r}$ and ${\bf r'}$ lie in the same angular direction. The radial difference $r-r'$, however, does not have to be small.

For simplicity we shall denote the scalar Green functions generically by $\Delta_l(r,r')$ (thus $\Delta_l$ is either $F_l$ or $G_l$). Their governing equation is
\begin{equation}
\left[ \frac{\partial^2}{\partial r^2}+\frac{2}{r}\frac{\partial}{\partial r}+\omega^2-\frac{l(l+1)}{r^2} \right]\Delta_l(r,r')
=-\frac{1}{r^2}\delta(r-r').
\end{equation}
\label{49}
The solution contains spherical Bessel functions, and has the general form
\begin{equation}
\Delta_l(r,r')=\frac{ik}{1-\tilde{C}_I\tilde{C}_{II}}\left[ j_l(kr_<)-\tilde{C}_Ih_l^{(1)}(kr_<)\right]
\left[h_l^{(1)}(kr_>)-\tilde{C}_{II}j_l(kr_>) \right],
\end{equation}
\label{50}
where $k=|\omega|$, $\tilde{C}_I ~ {\rm and}~ \tilde{C}_{II}$ being constants. This form satisfies the discontinuity condition following from Eq.~(49) on the radial derivative of $\Delta_l$ at $r=r'$, with the Wronskian $W\{j_l(x), h_l^{(1)}(x)\}=i/x^2$. Taking into account the boundary conditions at $r=a,b$ we can determine the constants: for the F (or TE) mode we get
\begin{equation}
\tilde{C}_{IF}(ka)=\frac{\tilde{s}_l(ka)}{\tilde{e}(ka)},~~~~
\tilde{C}_{IIF}(kb)=\frac{\tilde{e}_l(kb)}{\tilde{s}_l(kb)},
\end{equation}
\label{51}
whereas for the G (or TM) mode 
\begin{equation}
\tilde{C}_{IG}(ka)=\frac{\tilde{s}_l'(ka)}{\tilde{e}_l'(ka)},~~~~
\tilde{C}_{IIG}(kb)=\frac {\tilde{e}_l'(kb)}{\tilde{s}_l'(kb)}.
\end{equation}
\label{52}
Here, we let prime mean differentiation with respect to the whole argument.
We now perform a complex frequency rotation, $\omega \rightarrow i \hat{\omega},~k\rightarrow i|\hat{\omega}|=i\hat{k}$, and replace the conventional Riccati-Bessel functions $\tilde{s}_l(x)=xj_l(x),~\tilde{e}_l(x)=xh_l^{(1)}(x)$ by new ones $s_l,e_l$ defined according to
\[ s_l(x)=(-i)^{l+1}\tilde{s}_l(ix)=\sqrt{\frac{\pi x}{2}}\,I_{\nu}(x), \]
\begin{equation}
e_l(x)=i^{l+1}\tilde{e}_l(ix)=\sqrt{\frac{2x}{\pi}}\,K_{\nu}(x).
\end{equation}
\label{53}
Here $\nu=l+1/2,~I_\nu$ and $K_\nu$ are modified Bessel functions, and the Wronskian of importance now is $W\{s_l, e_l\}=-1$. The frequency rotation implies that we replace the "tilde" constants $\tilde{C}$ by new constants $C$, in accordance with the relation $\tilde{C}(ix)=(-1)^{l+1} C(x)$. Explicitly,
\[ C_{IF}(x)=\frac{s_l(x)}{e_l(x)},~~~~C_{IIF}(y)=\frac{e_l(y)}{s_l(y)}, \]
\begin{equation}
C_{IG}(x)=\frac{s_l'(x)}{e_l'(x)},~~~~C_{IIG}(y)=\frac{e_l'(y)}{s_l'(y)},
\end{equation}
\label{54}
where we here and henceforth let $x$ and $y$  be defined by $x=\hat{k}a,~y=\hat{k}b$. 

We now return to the expansions (45)-(48) for the spectral two-point functions.  Of main interest for us are these functions when the points ${\bf r}$ and ${\bf r'}$ are close to each other, but not overlapping. We moreover  set the time-splitting parameter $\tau=t-t'$ equal to zero. Substituting the two-point functions in Maxwell's stress tensor we can calculate the surface force density on either of the two surfaces. We choose the outer  surface $r=b$, since it will then become easy to relate the force to the free energy. Writing for simplicity $\langle E_r^2(r) \rangle$ instead of $\langle E_r(r)E_r(r') \rangle_{r' \rightarrow r}$, we obtain for $T=0$  the various two-point functions at $r=b-$: 
\begin{equation}
\langle E_r^2(b-) \rangle=\frac{(4\pi)}{\pi b^4} \int_0^{\infty}\frac{dy}{y}\sum_{l=1}^{\infty}
\frac{2l+1}{4\pi}l(l+1)\frac{s_l(y)-C_{IG}(x)e_l(y)}{s_l'(y)-C_{IG}(x)e_l'(y)},
\end{equation}
\label{55}
\begin{equation}
\langle E_{\perp}^2(b-) \rangle=\langle H_r^2(b-) \rangle =0,
\end{equation}
\label{56}
\begin{equation}
\langle H_{\perp}^2(b-) \rangle 
=\frac{-(4\pi)}{\pi b^4}\int_0^{\infty} ydy \sum_{l=1}^{\infty}
\frac{2l+1}{4\pi}[ \frac{s_l(y)-C_{IG}(x)e_l(y)}{s_l'(y)-C_{IG}(x)e_l'(y)}
+ \frac{s_l'(y)-C_{IF}(x)e_l'(y)}{s_l(y)-C_{IF}(x)e_l(y)} ]
\end{equation}
\label{57}
(the prefactors $4\pi$ again reflecting the Gaussian units).

Using Maxwell's stress tensor we can write the surface force density on the outer surface as
\begin{equation}
f_b=-\frac{1}{8\pi}\langle E_r^2(b-)\rangle +\frac{1}{8\pi}\langle H_{\perp}^2(b-)\rangle.
\end{equation}
\label{58}
Substituting Eqs.~(55) and (57) into Eq.~(58) we obtain, when taking into account the governing equation for the Riccati-Bessel functions,
\begin{equation}
s_l''(y)=(1+l(l+1)/y^2)s_l(y)
\end{equation}
\label{59}
(and similarly for $e_l(y)$), that
\begin{equation}
f_b=\frac{-1}{2\pi b^4}\int_0 ^{\infty}ydy \sum_{l=1}^{\infty}\frac{2l+1}{4\pi}
[ \frac{s_l'(y)-C_{IF}(x)e_l'(y)}{s_l(y)-C_{IF}(x)e_l(y)}
+\frac{s_l''(y)-C_{IG}(x)e_l''(y)}{s_l'(y)-C_{IG}(x)e_l'(y)} ].
\end{equation}
\label{60}
From this expression it is apparent how both modes F and G contribute to the force. 

We can write Eq.~(60) as a sum of the following two terms:
\begin{equation}
f_b=f_b^{(0)}+f_b^{int},
\end{equation}
\label{61}
where
\begin{equation}
f_b^{(0)}=\frac{-1}{2\pi b^4}\int_0^{\infty}ydy\sum_{l=1}^{\infty}\frac{2l+1}{4\pi}
[ \frac{s_l'(y)}{s_l(y)}+\frac{s_l''(y)}{s_l'(y)} ],
\end{equation}
\label{62}
\begin{equation}
f_b^{int}=\frac{-1}{2\pi b^2}\int_0^{\infty}d\hat{k}\sum_{l=1}^{\infty}\frac{2l+1}{4\pi}
\frac{\partial}{\partial b}\ln \{ [1-\frac{s_l(x)}{e_l(x)}\frac{e_l(y)}{s_l(y)}]
[1-\frac{s_l'(x)}{e_l'(x)}\frac{e_l'(y)}{s_l'(y)}] \},
\end{equation}
\label{63}
with $y=\hat{k}b$ (the operator $\partial /\partial b$ is taken at constant value of $a$). The expression (62) is the same as the inner contribution to the surface force on a perfectly conducting shell \cite{milton78, brevik94}. This term does not involve the interaction between the two media, and will be discarded in the following. Of interest for us is the interaction term (63). As $s_l(y)=\frac{1}{2}e^y$ and $e_l(y)=e^{-y}$ for large $y$, it is evident from (63) that $f_b \rightarrow 0$ if the outer surface recedes to infinity while the inner surface is kept constant. This is physically as it should be.

A remark is here in order, concerning the physical meaning of the two force terms in Eq.~(61), in particular the possibility of making measurements. It ought to be emphasized, first of all, that the term $f_b^{(0)}$ is a mathematical construct. It does not seem to be possible to measure this term, not even in principle. If one imagines the case of a perfectly conducting singular shell with radius $a$, the case studied in \cite{milton78} and also in \cite{brevik94}, then $f_b^{(0)}$ has to be taken together with a similar term on the {\it outside} to make up a surface force that in principle ought to be measurable. However, as far as we know, no measurement has so far been made for even such a complete shell. In our case, what {\it is} measurable, at least in principle, is the interaction term (63). For instance, one might envisage to measure the attractive force between a micrometer-sized conducting sphere and a semi-spherical trough in a conducting plate, thus some kind of generalization of the atomic force microscope measurements reported in \cite{mohideen98} and \cite{roy99}.  

The expressions above refer to zero temperature. The transition to finite temperatures is made by means of a discretization of the frequencies,
\begin{equation}
\hat{k} \rightarrow K=2\pi n/\beta,~~~~x\rightarrow Ka,
\end{equation}
\label{64}
with $n$ an integer. The rule for going from frequency integral to a sum over Matsubara frequencies is
\begin{equation}
\int_0^{\infty}d\hat{k} \rightarrow \frac{2\pi}{\beta} {\sum_{n=0}^\infty}\, ',
\end{equation}
\label{65}
where the prime on the summation sign means that the $n=0$ term is taken with half weight. The finite-temperature force expression accordingly becomes (recall that $\nu=l+1/2$)
\begin{equation}
f_b^{int}=\frac{-1}{2\pi b^2\beta}{\sum_{n=0}^\infty }\, ' \sum_{l=1}^\infty \nu\frac{\partial}{\partial b}
\ln \{ [1-\frac{s_l(x)}{e_l(x)}\frac{e_l(y)}{s_l(y)}]
[1-\frac{s_l'(x)}{e_l'(x)}\frac{e_l'(y)}{s_l'(y)}] \},
\end{equation}
\label{66}
where now $x=2\pi na/\beta$,  $y=2\pi nb/\beta$.

As for measurements of a force like $f_b^{int}$ there are subtle problems although the free energy (Eq.~(68) below) is well defined. One may imagine that the two spherical media are liquids, and that the outer spherical shell can move. (The inner shell can also move if liquid is added or removed through a small pipe.) However, there is an extra complication compared to the case of parallel plates, namely the change of {\it radius} of a spherical surface. This will change the free energy associated with the surface tension. Although the latter change of energy is finite, a more precise evaluation of it will obviously be a complex task. The molecular structure has to be taken into account at the surface where the density changes abruptly.

It can be noted, however, that the sum of the two surface energies must be the opposite of the general result (40) for $a=b$, i.~e., with the two spherical surfaces fused together. One is then left with no surfaces at all, and the bulk free energy applies everywhere. However, simply putting $a=b$ in a continuum approach does not work; Eq.~(40) will diverge unless a cutoff is introduced for large $l$ to mimic the molecular diameter. Whether the self-force $f_b^{(0)}$ in Eq.~(62) is possibly related to such a surface tension energy or not is a problem that we have so far not been able to clear up. 

\section{The free energy, analytically and numerically}

Since we have calculated the force density on the {\it outer} surface due to the mutual interaction, it is easy to derive the corresponding expression for the interaction free energy $F^{int}$. We imagine the outer surface to be displaced by a small amount $db$, while the inner surface is kept constant. The relation 
\begin{equation}
f_b^{int}=-\frac{1}{4\pi b^2}\frac{\partial F^{int}}{\partial b}
\end{equation}
\label{67}
is integrated from $b$ to infinity, noting that $F^{int}=0$ at $b=\infty$. Making use of (66) we thus obtain, dropping the superscript 'int',
\begin{equation}
\beta F=2{\sum_{n=0}^\infty}\, ' \sum_{l=1}^\infty  \nu
\ln \{ [1-\frac{s_l(x)}{e_l(x)}\frac{e_l(y)}{s_l(y)}]
[1-\frac{s_l'(x)}{e_l'(x)}\frac{e_l'(y)}{s_l'(y)}] \},
\end{equation}
\label{68}
valid at arbitrary temperatures. Comparing with the more general result of Eqs.~(38)-(40) in Sect. 5 one finds that Eq.~(68) agrees with these when $\varepsilon \rightarrow \infty$ as then $\varepsilon s_a' s_{a\varepsilon} \gg s_a s_{a\varepsilon}',~~s_a's_{a\varepsilon} \ll s_a s_{a\varepsilon}'$, etc. (with $s_a=s_l(x)$ etc. as introduced above Eq.~(32)).

At $T=0$, where the free energy $F$ is the same as the energy $E$, we can write
\begin{equation}
E=\frac{1}{\pi a}\int_0^\infty dx \sum_{l=1}^\infty \nu \ln \{ [1-\frac{s_l(x)}{e_l(x)}\frac{e_l(y)}{s_l(y)}]
[1-\frac{s_l'(x)}{e_l'(x)}\frac{e_l'(y)}{s_l'(y)}] \}.
\end{equation}
\label{69}

Expressions (68) and (69) hold for arbitrary widths of the annular region. It is of interest, before embarking on numerical evaluations, to analyse some limiting cases by analytical means. The limiting case of immediate interest is that of a narrow slit, i.e.,  
\begin{equation}
\xi \equiv \frac{b-a}{a} \ll 1.
\end{equation}
\label{70}
This case is motivated physically from the fact that the Casimir measurements are made for small separations only, and also because we have in this way the possibility to check our results against the standard results for parallel plates in the limit when $\xi \rightarrow 0$.

At $T=0$ we find, when $x$ and $y$ lie close to each other \cite{brevik94},
\begin{equation}
\frac{s_l(x)}{e_l(x)}\frac{e_l(y)}{s_l(y)}=\frac{s_l'(x)}{e_l'(x)}\frac {e_l'(y)}{s_l'(y)}
=e^{-\nu \phi},
\end{equation}
\label{71}
to $ O(1/\nu )$ in the uniform asymptotic (or Debye) expansion. Here
\begin{equation}
\phi=2\xi \sqrt{1+z^2}[ 1-\frac{1}{2}\frac{\xi}{1+z^2}+...],~~~z=\frac{x}{\nu}.
\end{equation}
\label{72}
Keeping only the first term, we find the $T=0$ interaction energy to be
\begin{equation}
E=\frac{2}{\pi a}\sum_{l=1}^{\infty}\nu^2 \int_0^{\infty}dz \,\ln ( 1-e^{-2\xi\nu\sqrt{1+z^2}} ).
\end{equation}
\label{73}
This expression can be processed further, if we make a power expansion of the logarithm and take into account the property $\sum_{l=1}^{\infty} \nu^2 e^{-\nu \phi} \rightarrow 2/\phi^3$ when $\phi \rightarrow 0$ \cite{brevik83}. Then,
\begin{equation}
E=\frac{-1}{2\pi a\xi^3}\zeta (4)\int_0^{\infty} \frac{dz}{(1+z^2)^{3/2}}=
-\frac{\pi^3}{180a}\frac{1}{\xi^3},
\end{equation}
\label{74}
which corresponds to the following interaction energy per unit surface (total surface area $A=4\pi a^2$, and $d=b-a$):
\begin{equation}
\frac{E}{A}=-\frac{\pi^2}{720}\frac{1}{d^3}.
\end{equation}
\label{75}
This simple calculation thus provides us with a satisfactory check: Eq.~(75) is the  conventional expression for the Casimir energy of two parallel plates. (The expression includes the effect of retardation. That is, the distance $d$ is much larger than the characteristic wavelength of the absorption spectrum of the medium.) Thus, by retaining the first order term in $\xi$ we see that our theory reduces to the standard $T=0$ theory. Corrections to the theory arising from the curvature of the surfaces can in principle be worked out by going to larger powers in $\xi$.

At {\it finite temperatures}, we obtain for the free energy
\begin{equation}
\beta F=4{\sum_{n=0}^{\infty}}\,' \sum_{l=1}^{\infty} \nu\ln (1-e^{-2\xi\sqrt{\nu^2+n^2t^2}}),
~~~~t=2\pi a/\beta.
\end{equation}
\label{76}
This expression, as before, implies keeping of only the first term in the expansion (72), but it puts no restriction on the temperature. 

Let us consider the limiting case of high temperatures, first going back to the expression (68), holding for arbitrary widths $d$. For the highest temperatures (classical limit), only the lowest Matsubara frequency ($n=0$) contributes. As $x=nt,~~y=bnt/a$, it is seen that we then need to evaluate $s_l$ and $e_l$ when the  arguments tend to zero. As
\begin{equation}
s_l(x)=\frac{\sqrt{\pi}}{\Gamma(\nu+1)} ( \frac{x}{2} )^{(\nu+\frac{1}{2})},~~~
e_l(x)=\frac{\Gamma(\nu)}{\sqrt{\pi}} ( \frac{x}{2} )^{(-\nu+\frac{1}{2})}
\end{equation}
\label{77}
for small arguments, we have
\begin{equation}
\frac{s_l(x)}{e_l(x)}\frac{e_l(y)}{s_l(y)}=\frac{s_l'(x)}{e_l'(x)}\frac{e_l'(y)}{s_l'(y)}
=(\frac{a}{b})^{2l+1},
\end{equation}
\label{78}
so that the contribution from $n=0$ becomes
\begin{equation}
\beta F(n=0)=\sum_{l=1}^{\infty}(2l+1)\ln [1-(\frac{a}{b})^{2l+1} ].
\end{equation}
\label{79}
This is in agreement with our previus expression (18) ($A_l=1$ when $\varepsilon \rightarrow \infty$), except from a factor 2. The physical reason for this is that both F and G modes contribute to Eq.~(79), whereas only one mode contributes in (18). This artifact in Eq.~(79) is related to the fact that $\varepsilon = \infty$ is considered, while in Eq.~(18) $\omega =0$ is considered {\it before} the limit $\varepsilon \rightarrow \infty$ is taken.

In the case of a narrow slit we obtain from (76) the $n=0$ contribution
\begin{equation}
\beta F(n=0)=2\sum_{l=1}^{\infty}\nu\ln(1-e^{-2\xi\nu}).
\end{equation}
\label{80}
Again making a power expansion of the logarithm, and observing the relationship $\sum_{l=1}^{\infty}\nu e^{-\nu \phi}=1/\phi^2$ when $\phi\rightarrow 0$, we get
\begin{equation}
\beta F(n=0)=-\frac{\zeta(3)}{2}\frac{1}{\xi^2},
\end{equation}
\label{81}
which corresponds to
\begin{equation}
\frac{\beta F(n=0)}{A}=-\frac{\zeta(3)}{8\pi}\frac{1}{d^2}.
\end{equation}
\label{82}
Again, this is a satisfactory check, as Eq.~(82) is the conventional high-temperature result for parallel plates.

For a {\it narrow slit} we may also obtain the known result for a parallel plates configuration more generally. In the wave equation the term $l(l+1)/r^2 \simeq l(l+1)/a^2$ is replaceable with $k_\perp ^2$ where ${\bf k}_\perp $ is the transverse wave vector, i.e.,
\begin{equation}
\nu^2 =(l+\frac{1}{2})^2 \simeq l(l+1)= k_\perp^2 a^2.
\end{equation}
\label{83}
When $l$  is large we can regard it as continuous quantity, whereby the sum can be replaced by an integral. We have then
\begin{equation}
\sum_{l=1}^\infty (2l+1) \rightarrow \int(2l+1)dl=2a^2\int k_\perp \,dk_\perp.
\end{equation}
\label{84}
Further,
\[ \xi nt=\frac{2\pi n}{\beta}d=Kd, \]
\begin{equation}
\xi \nu=\frac{d}{a}k_\perp a=k_\perp d,
\end{equation}
\label{85}
or
\begin{equation}
2\xi \sqrt{\nu^2+n^2t^2}=2qd,~~~\mbox{with}~~~q^2=k_\perp^2+K^2.
\end{equation}
\label{86}
Insertion of this into Eq.~(76) yields
\begin{equation}
\beta F=4a^2{\sum_{n=0}^\infty}\,'\int_{\zeta_n}^\infty qdq\,\ln(1-e^{-2qd}),
\end{equation}
\label{87}
using $ qdq=k_\perp dk_\perp $. The surface force density thus becomes ($b \rightarrow a$)
\begin{equation}
f_b=-\frac{1}{4\pi a^2}\frac{\partial F}{\partial d}=-\frac{2}{\pi \beta}
{\sum_{n=0}^\infty}\,'\int_{\zeta_n}^\infty q^2dq\,\frac{e^{-2qd}}{(1-e^{-2qd})},
\end{equation}
\label{88}
with $\zeta_n=K=2\pi n/\beta$ being the Matsubara frequency. This is in agreement with Eq.~(2.9) in \cite{hoye98} ($A_n=B_n=1$ for $\varepsilon = \infty$). It is also in agreement with Eq.~(3.8) in \cite{schwinger78} (it should be mentioned that $q^2$ in our present notation, Eq.~(86), is the same as $\kappa^2$ in \cite{schwinger78}, and also that the distance $d$ above is the same as $a$ in \cite{schwinger78} and \cite{hoye98}). Note that whereas the expression (88) presupposes a narrow slit (large $l$), there is no restriction on the temperature.

After these preliminary analytic considerations we now present numerically calculated results for the free energy, when the walls are conducting ($\varepsilon = \infty$).
 All results are given in nondimensional form. The numerics turned out to be rather demanding; as preliminary tests indicated that Matlab would be insufficient for our purpose we turned to standard FORTRAN routines and made use of them throughout. On a logarithmic plot with base 10, 
Fig.1 shows how $\lg(-\beta Ft)=\lg(-2\pi aF)$ varies with relative width $d/a$ for various values of the nondimensional temperature $t=2\pi a/\beta$. At zero temperature, an integration routine was used for the $x$ integral in Eq.~(69). At finite temperatures, the double sum in Eq.~(68) was calculated as it stands (thus without expansion procedures for the Riccati-Bessel functions), with  use of the FORTRAN library for the Bessel functions of half-integer order. Allowing the numerical tolerances in the $l$ sum as well as in the $n$ sum to be equal to $10^{-6}$, we found for the case of $t=0.01$ and $d/a=0.1$ the necessary number of terms in the $n$ sum to be about 415 000. For larger widths the necessary number of terms turned out to be considerably less; for instance for $d/a=1$ and the same $t$ the number was about 8900. 

A characteristic property of the curves in Fig.1 is that they tend to overlap in the case of {\it low} temperatures. Thus the curve calculated for $t=0$ (via an $x$ integral and an $l$ sum) is indistinguishable from other curves calculated in the whole temperature region (via a double sum) up to $t=1$. Numerically, for $d/a=0.1$ the difference in $\lg (-\beta Ft)$ between the cases $t=0$ and $t=1$ is found to be about $10^{-4}$.
 For zero temperature, as mentioned, $F$ is the same as the energy $E$.

Whereas the representation in Fig.1 is most useful for low temperatures, we show in Fig.2 the representation of $\lg (-\beta F)$ versus $d/a$. This is convenient for high temperatures, since now the curves for high $t$ stay inside the figure and tend to overlap. The curve calculated for $t=50$ is actually indistinguishable from the shown curve referring to $t=200$. 
This fact reflects that $F$ is proportional to $t$, thus in accordance with the behaviour of 
high temperature mutual free energies for classical harmonic oscillators in general. For a narrow slit, we expect that the calculation agrees with the approximate formula (82). We may check this in the case of $t=200$, $d/a=0.05$: the machine calculation then yields $\beta F=-249.7$, whereas Eq.~(82) yields $\beta F=-\frac{1}{2}\zeta (3)(a/d)^2= -240.4$, thus an error of 4 \%.  The reason why the accuray here is only moderate, is most likely that the width parameter $d/a=0.05$ is not small enough to represent a narrow slit to a high precision. Generally, it turned out to be difficult to calculate cases of higher temperatures or more narrow slits than those shown in the figures, without entering into special alterations of the FORTRAN routines.

Finally, it is of interest to display explicitly the free energy's low-temperature plateau, and its high-temperature proportionality to $t$, for a fixed value of the relative width. This is done in Fig.3, for the cases of $d/a=\{0.05, 0.075, 0.1\}$. The horisontal plateau is seen to prevail quite accurately until a gradual increase takes place in the region roughly between $t=10$ and $t=30$ ($\lg 30=1.477$). It is notworthy that this behaviour is in agreement with the following simple physical argument: The most significant frequencies $\omega$ contributing to $F$ are generally of the same order as the inverse width of the gap, {\i.e.} $\omega \sim 1/d$. Now, from Wien's displacement law we know that the maximum of a blackbody distribution occurs at a frequency of $\omega_m =2.8/\beta$. Temperature effects are expected to become significant  when $\omega$ is comparable to $\omega_m$. Putting $\omega \sim \omega_m$ we obtain $t=2\pi a/\beta \sim 2a/d$. In the present case this amounts to $t \sim 20 - 40$,  ($\lg t \sim 1.3 - 1.6$), which is seen to be in good  agreement with the location of the shoulder in Fig.~3. Moreover, for higher temperatures, the proportionality of $F$ to $t$ is clear from the figure.

\bigskip

\bigskip

{\bf Acknowledgement}

\bigskip

We thank Gabriel Barton for sending us a copy of his recent preprint \cite{barton00}. The reader is referred also to the finite-temperature paper of Klich {\it et al.} \cite{klich00}.

\newpage
{\bf Figure Captions}

\bigskip 

{\bf Figure 1} 

 Logarithm (base 10) of mutual nondimensional free energy, $ \lg (-\beta Ft)=\lg (-2\pi aF)$, versus relative  width $d/a$ for various values of the nondimensional temperature $t=2\pi a/\beta$. For $0 \leq t<1$ the curves are overlapping (i.e., only one curve is drawn), consistent with Fig.3.

\bigskip

{\bf Figure 2}

 Same as Fig.1, but $\lg (-\beta F)$ is shown, as appropriate for the case of high temperatures. For $t$ larger than about 50 the curves are overlapping (only one curve is drawn), consistent with Fig.3.

\bigskip

{\bf Figure 3}

 Variation of $F$ versus $t$ for the case of $d/a=\{0.05, 0.075, 0.1\}$.

\end{document}